\documentclass[aps,prb,twocolumn,epsf,epsfig]{revtex4}
\usepackage{graphicx} \usepackage{amssymb,amsmath, amsthm}

\newcommand{\beq}{\begin{equation}} 
\newcommand{\eeq}{\end{equation}}
\newcommand{\barray}{\begin{eqnarray}}

\begin{document}


%
\title{Finite temperature properties of the triangular lattice $t$-$J$ model, applications to Na$_x$CoO$_2$}


\author{Jan O. Haerter, Michael R. Peterson, B. Sriram Shastry}
\address{Physics Department, University of California,  Santa Cruz, Ca
95064 } \date{\today}




\begin{abstract}
We present a finite temperature ($T$) study of the $t$-$J$ model on
the two-dimensional triangular lattice for the negative hopping $t$,
as relevant for the electron-doped Na$_x$CoO$_2$ (NCO). To understand
several aspects of this system, we study the $T$-dependent chemical
potential, specific heat, magnetic susceptibility, and the dynamic
Hall-coefficient across the entire doping range. We show
systematically, how this simplest model for strongly correlated
electrons describes a crossover as function of doping ($x$) from a
Pauli-like weakly spin-correlated metal close to the band-limit
(density $n=2$) to the Curie-Weiss metallic phase ($1.5<n<1.75$) with
pronounced anti-ferromagnetic (AFM) correlations at low temperatures
and Curie-Weiss type behavior in the high-temperature regime. Upon
further reduction of the doping, a new energy scale, dominated by
spin-interactions ($J$) emerges (apparent both in specific heat and
susceptibility) and we identify an effective interaction $J_{eff}(x)$,
valid across the entire doping range. This is distinct from Anderson's
formula\cite{NT_weakening}, as we choose here $t<0$, hence
the opposite sign of the usual Nagaoka-ferromagnetic situation. This
expression includes the subtle effect of weak kinetic
AFM\cite{counter_nagaoka} - as encountered in the infinitely
correlated situation ($U=\infty$). By explicit computation of the
Kubo-formulae, we address the question of practical relevance of the
high-frequency expression for the Hall coefficient
$R_H^*$\cite{sss}. We hope to clarify some open questions concerning
the applicability of the $t$-$J$ model to real experimental situations
through this study.


\end{abstract}

 \maketitle


\section{Introduction}

Since the discovery of superconductivity in H$_2$O-intercalated NCO
crystals \cite{takada} this compound has drawn significant
attention. In addition, NCO displays a variety of remarkable
thermoelectric effects, such as an unusually large and magnetic field
dependent thermopower\cite{ong_nature} and a $T$-linear
Hall-coefficient\cite{wang_hall, sss}. While some of its properties
are similar to those seen in high-$T_c$ materials there are some
distinctive differences. Both the cobaltates and the cuprates become
superconducting in the vicinity of the Mott-insulating state, however,
while this is achieved through hole-doping in the cuprates, NCO
requires electron-doping $x$ between $x=1/4$ and $x=1/3$
\cite{foo}. While for both compounds the role of electron transport in
two-dimensional planes - separated by thick insulating layers -
appears to be crucial, the sites for the localized electron
wavefunctions are organized in planar square lattices (Cu-ions) in the
cuprates and in a triangular lattice (Co-atoms) in NCO. The most
striking differences however, are the sign of the hopping, which is
inferred to be positive in the cuprates and negative in NCO, and its
absolute value, which is believed to be an order of magnitude smaller
in NCO\cite{hasan}. While the experimental results indicate a rich
blend of phases - stemming perhaps from structural rearrangements with
$x$ - it is first necessary to systematically probe the scope of a
single model for strongly correlated systems - namely the $t$-$J$
model - in the case of this electronically frustrated triangular
lattice.

In the half-filled band, where strong interactions lead to the
spin-$\frac{1}{2}$ Heisenberg-antiferromagnet (HAF), the square
geometry facilitates anti-ferromagnetism (AFM) through its bipartite
nature, while the tendency towards AFM in the cobaltate system is
weaker. However, extensive numerical computations \cite{bernu}
indicate N\'eel long-ranged order even in the triangular lattice. Away
from half-filling, projected hopping on the square lattice system
($t>0$) obeys the Nagaoka mechanism \cite{nagaoka}. This results in a
competition between the ferromagnetic (FM) kinetic operator and AFM
potential energy for finite values of $J$. Conversely, in the current
case of the triangular lattice with a negative hopping $t<0$, AFM
tendencies are {\it supported} - rather than weakened - by the
electronically frustrated kinetic energy operator. By further reducing
the $t$-$J$ model - taking $J=0$ - we explicitly isolate the effect of
electronic frustration which leads to {\it kinetic Antiferromagnetism
(kA)}\cite{counter_nagaoka}, an effect triggered only by the motion of
holes in an AFM spin-background, even away from the special case of
doping $1/L$, with $L$ the number of sites.

This paper has two main objectives: Firstly, we want to supplement the
arguments made in ref. \onlinecite{curie_weiss}, where the Curie-Weiss
phase in the doping range $x>0.5$ was addressed. Secondly, we will
discuss the finite temperature $t$-$J$ model starting from the
Nagaoka-regime and then extend into finite dopings. The structure is
as follows: In section II we describe our model and the exact
diagonalization technique, which we utilize for our
computations. Section III presents the results for the static
properties chemical potential, specific heat and entropy as well as
the spin susceptibility. We then investigate the dynamic
Hall-coefficient $R_H(\omega)$ and compare its dc- and high-frequency
behavior. Lastly, we draw a connection between the Hall-coefficient
and the diamagnetic susceptibility. For all properties, we study
systematically the dependence on temperature, doping, and interaction
strength. In Section IV we conclude and summarize our 
results.

\setlength{\unitlength}{1cm}
\begin{center}
\begin{figure*}
\begin{tabular}[t]{cc}
\includegraphics[width=6.4cm]{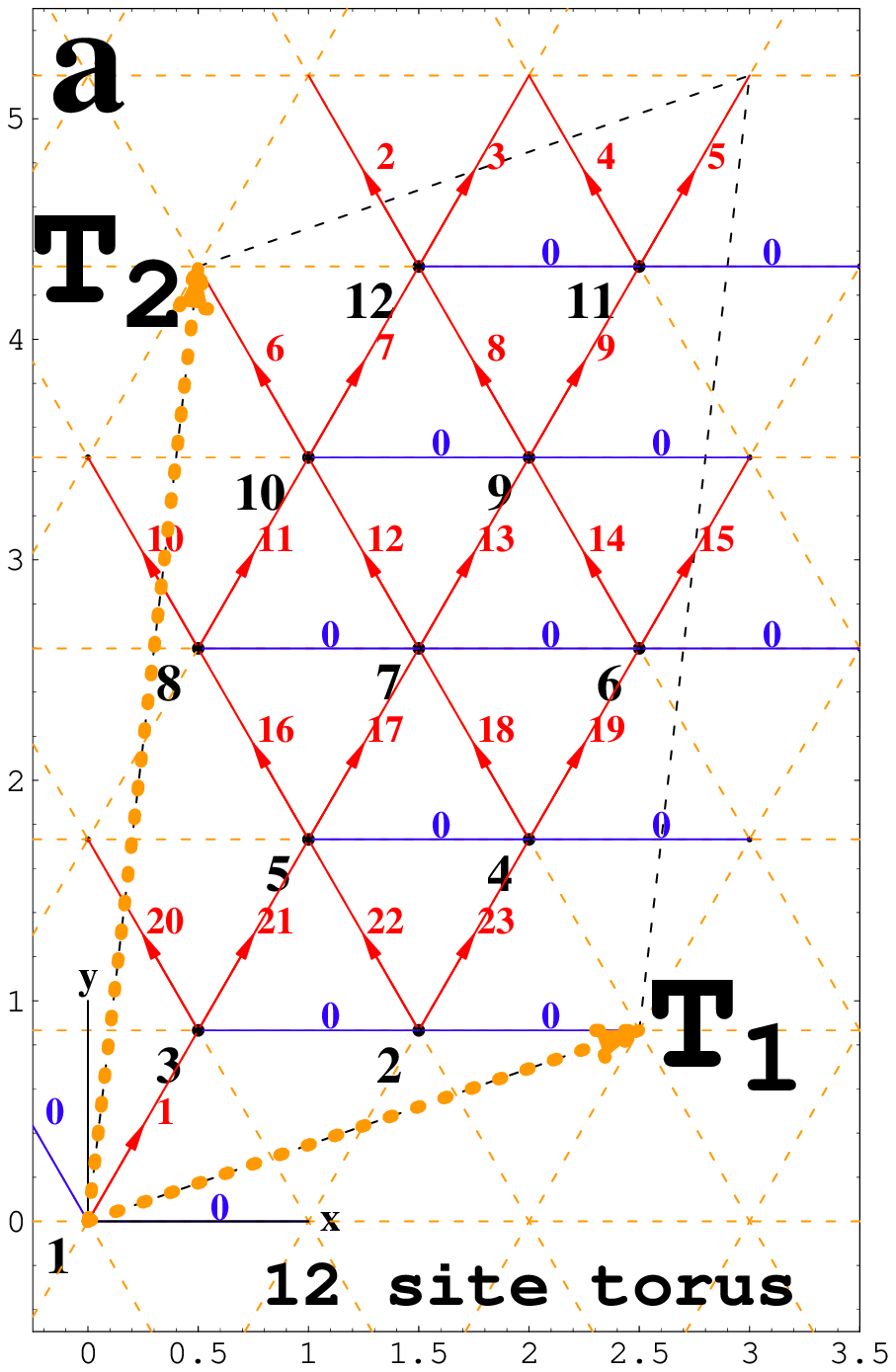}
\begin{tabular}[b]{c}
\includegraphics[width=8.0cm]{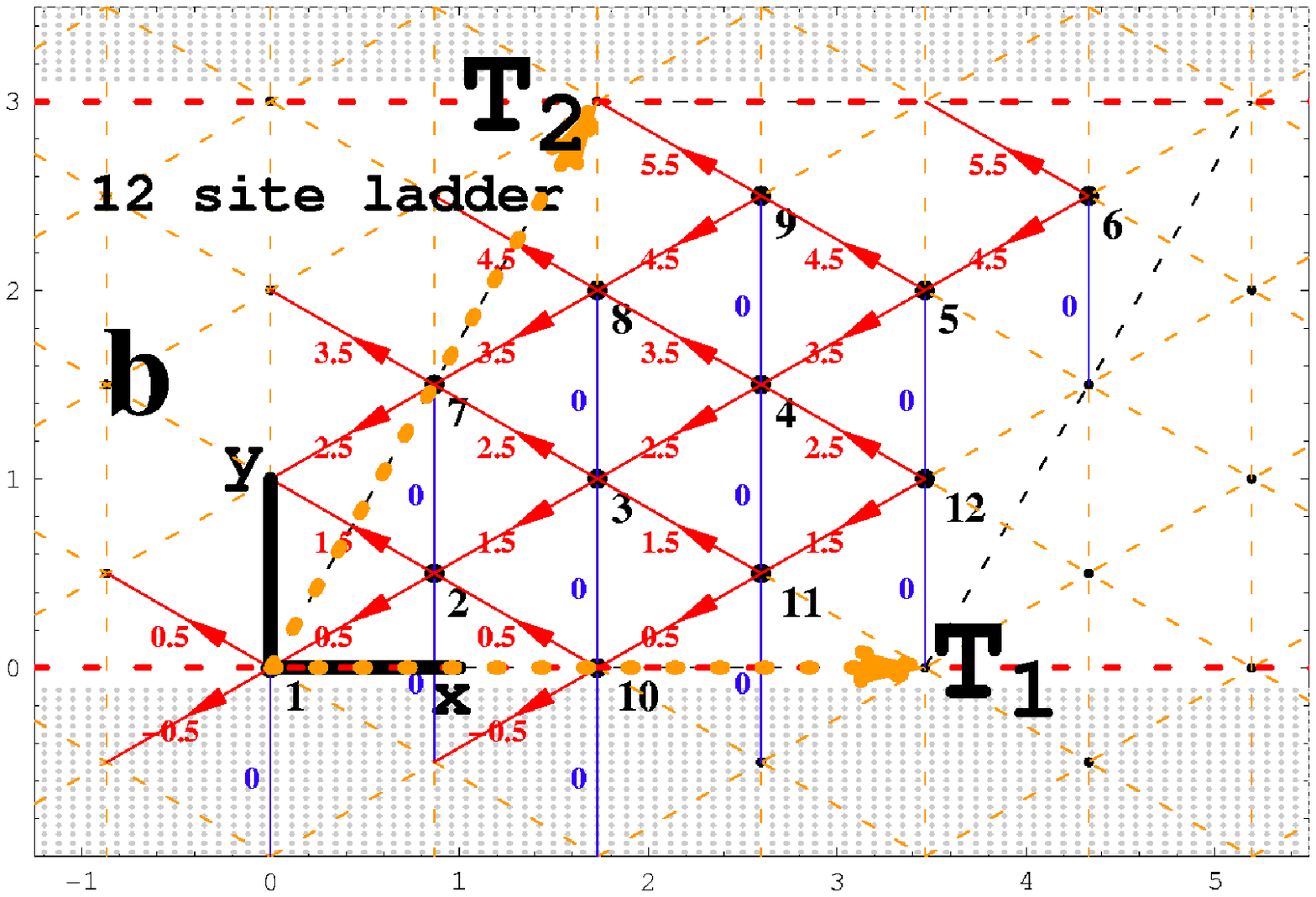}\\
\includegraphics[width=4.5cm]{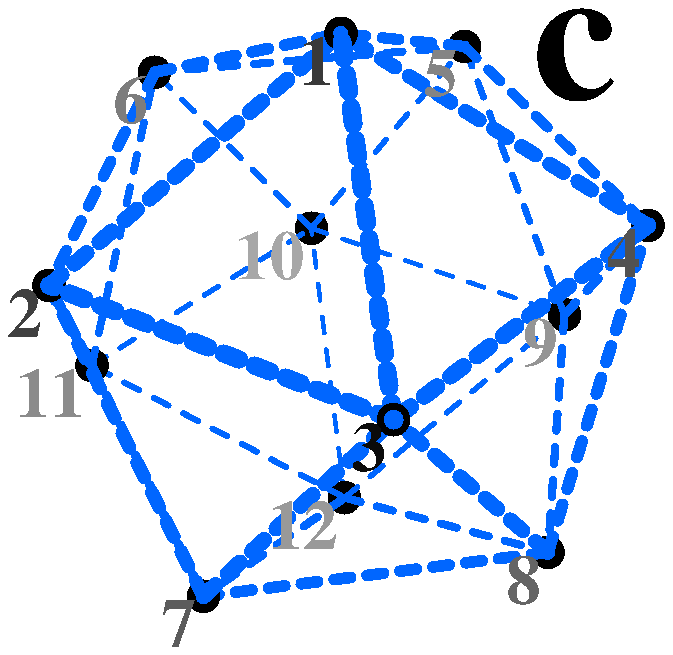}
\end{tabular}
\end{tabular}
\caption{ (color on-line) Small numbers along arrows (red) denote
Peierls phases in units of the flux through a single plaquette, phases
on Icosahedron (c) not shown for clarity. Dotted parallelograms
(black) specify boundaries of finite clusters ((a) and (b)). All sites
on corners of parallelograms are equivalent and translations along
bounding vectors ${\bf T_1}$ and ${\bf T_2}$ tile infinite lattice;
shaded region in (b) forbidden for ladder. Note: different coordinate
systems are used for torus and ladder. For ladder-system we use the
Landau gauge with ${\bf A} \propto y\hat{x}$, (a) represents only one
choice (denoted Torus a) for 12 sites. Another (denoted Torus b)
results from ladder (b) by connecting all corners through ${\bf T_1}$
and ${\bf T_2}$ (not shown). Torus c is simple $3\times 4$ cluster
(not shown).}
\label{cluster_example}
\end{figure*}
\end{center}

Beginning at a doping of only a single hole in the infinitely
correlated system, we track the effect of {\it kA} into the
intermediate doping regime. The specific heat develops a
low-temperature peak indicative of effective spin-interactions,
generated by a dressed hole in an AFM-spin background. The same effect
is manifest in the susceptibility results where we present an $x$- and
$J$-dependent effective interaction, with a small but finite AFM value
in the case of $J=0$. We proceed by an explicit computation of the
frequency-dependent Hall-coefficient. The results show weak
frequency-dependence of $R_H$ and encourage the applicability of its
high-frequency limit $R_H^*$ to practical situations. While small in
magnitude compared to the spin-susceptibility, we show through an
identity in terms of derivatives - suggested by our numerical results
- that the diamagnetic susceptibility alone may be closely connected
to the temperature dependent Hall-coefficient.

\section{Model and exact diagonalization}
As a model for hole doped strongly interacting Mott-insulators, the
$t$-$J$ Hamiltonian can be applied to the experimental situation of
electron-doped NCO after an electron-hole transformation\cite{kumar},
making use of the symmetry of the Hubbard model in respect to the
Mott-insulator. The replacement rules are $t\rightarrow -t$, $\delta
=|1-n|=x$ and $q_e\rightarrow -q_e$, where $\delta$ is now hole-doping
away from half-filling and $n$ is the electron density per site. This
leads to the $t$-$J$ model  \beq \hat{H}=-t\sum_{<i,j>,\sigma}
\hat{P}_G \hat{c}_{i \sigma}^{\dagger}\hat{c}_{j
\sigma}\hat{P}_G+J\sum_{<i,j>}\left(\hat{S}_i \cdot
\hat{S}_j-\frac{\hat{n}_i \hat{n}_j}{4}\right)\eeq where $ \hat{c}_{i
\sigma}^{\dagger}$ ($\hat{c}_{i \sigma}$) creates (annihilates) an
electron of spin $\sigma$, $\hat{S}_i$ is the three-component
spin-operator, $\hat{n}_i$ is the number operator and $i$ specifies
the lattice site. $\hat{P}_G$ denotes the Gutzwiller projector and the
summation is over all nearest neighbor pairs $<i,j>$.

In the following, whenever doping $x$ is mentioned this refers to
electron doping in the NCO system and hole-doping in the model
Hamiltonian. To compute the complete spectrum - as required for
thermodynamic properties - we employ the canonical ensemble exact
diagonalization technique for small clusters. The computational demand
limits the analysis to quite small systems and finite size effects are
challenging to overcome. Nonetheless, careful comparison between
several systems of different geometry enables us to extract the stable
behavior and somewhat isolate finite size effects. Conversely, finite
size effects can help identify the favored eigenstate of a system.

We have employed several toroidal and ladder geometries with $L=$ $9$,
$12$, and $18$ sites, as well as the highly symmetric icosahedral
geometry. The icosahedron - a 12-sited Platonic solid with 20
equilateral triangular faces - shares the property of geometric
frustration with the infinite system while the effect of the reduced
coordination number (z=5) appears to have only a small, quantitative
effect on the computational results
(Fig. \ref{cluster_example}(c)). Our 12-site torus clusters are
depicted in Fig. \ref{cluster_example}(a)-(b). A third 12-site cluster
is the simple $4\times 3$-cluster (called Torus c, not pictured).

While the numerical effort of exact computations for translationally
invariant systems can be significantly reduced by exploiting space
group symmetries of the Hamiltonian, these symmetries are broken upon
the introduction of a magnetic field, as relevant for the evaluation
of the Hall coefficient. A magnetic field is introduced by the usual
Peierls substitution which modifies the hopping $t$ between sites $i$
and $j$ by  \beq\label{peierls} t\rightarrow t_{ij}({\bf A})=t\exp
\left( i\frac{2\pi}{\phi_0}\int_i^j {\bf A}\cdot d{\bf
s}\right)\;,\eeq where ${\bf A}$ is the magnetic vector potential and
$\phi_0$ the flux quantum. We define the flux threading a triangular
plaquette as $\alpha\equiv \frac{2\pi}{\phi_0}\oint_{\vartriangle}
{\bf A}\cdot d{\bf s}$. In finite systems, the value of the smallest
non-zero magnetic field is limited to values of $\alpha\geq \pi/l$
where $l$ is the length of a periodic loop in the system. Through a
particular gauge we can achieve $l$ equal to the number of triangular
faces in the cluster, this guarantees the equality of the flux values
through all plaquettes. An example is given in
Fig. \ref{cluster_example}(a). A similar strategy has been followed in
the case of the square lattice quantum Hall effect\cite{kohmoto}. We
complement the computations on periodic systems with ladder systems,
which enable an infinitesimal flux to be chosen due to open boundary
conditions in one of the spatial dimensions
(Fig. \ref{cluster_example}(b)) and the Landau-gauge is adequate. In
the case of the icosahedron (Fig. \ref{cluster_example}(c)) the
magnetic field is applied through a (fictitious) magnetic monopole in
its center, creating equal magnetic fluxes through all triangular
plaquettes.

\section{Results}
\subsection{Chemical Potential}
The chemical potential is evaluated as a derivative of the free energy
$F$ with respect to the particle number $N$, \[\mu(T)=\frac{\partial
F(T)}{\partial N}\;.\] The derivative has to be taken as a discrete
difference of $F(N)$. To take this derivative numerically, we use the
central difference formula. Hence it is suitable to define
\[\mu^+=F(N+1)-F(N) \;{ \verb and  }\; \mu^-=F(N)-F(N-1)\;.\] The
value of $\mu$ can then be obtained from \[\mu=\frac{(\mu^++\mu^-)}{2}\;.\] In
Fig. \ref{mu_J5}(a)-(b) we present $\mu(T)$ at several $J$ and dopings
$x$. At low temperatures $\mu(T)$ shows degenerate Fermi behavior
($\propto T^2$) \cite{gs_deg}. In the strong-correlation regime ($J=0$
and $J=0.4|t|$) the ground state (g.s.) internal energy has a maximum
at intermediate dopings $x\approx 0.5$ due to the competition of
increasing number of mobile particles and reduction of the available
vacancies upon decrease of doping and, hence, the low temperature
chemical potential changes sign here. In the high-temperature regime
the doping-dependence of the Hilbert-space dimension determines the
limit of \[\mu(T\rightarrow\infty)\propto T\frac{\partial
g(x)}{\partial x}\] with $g(x)=(1-x)\log (2x/(1-x))+\log (1/x)$ essentially the
natural logarithm of the dimension of the Hilbert-space. In the
projected $t$-$J$ model, $g(x)$ peaks at $x=1/3$. For $x<1/3$ the
slope at large $T$ becomes negative. In Fig. \ref{mu_J5} the
comparison of $\mu(T)$ for different interactions shows that the
impact of projected fermions near half-filling is to force a negative
slope of $\mu$ for $x>0$ which occurs in the non-interacting system
only for $x<0$. This is an example of the qualitative difference
between a strongly-interacting Fermi-system and its non-interacting
counterpart.

\setlength{\unitlength}{1cm}
\begin{center}
\begin{figure*}
\begin{picture}(0,0)(0,0)
\put(-6.9, -.7){{\large (a)}} \put(-3.3, -.7){{\large (b)}} \put(.3,
-.7){{\large (c)}} \put(3.9, -.7){{\large (d)}}
\end{picture} 
\includegraphics[width=18cm]{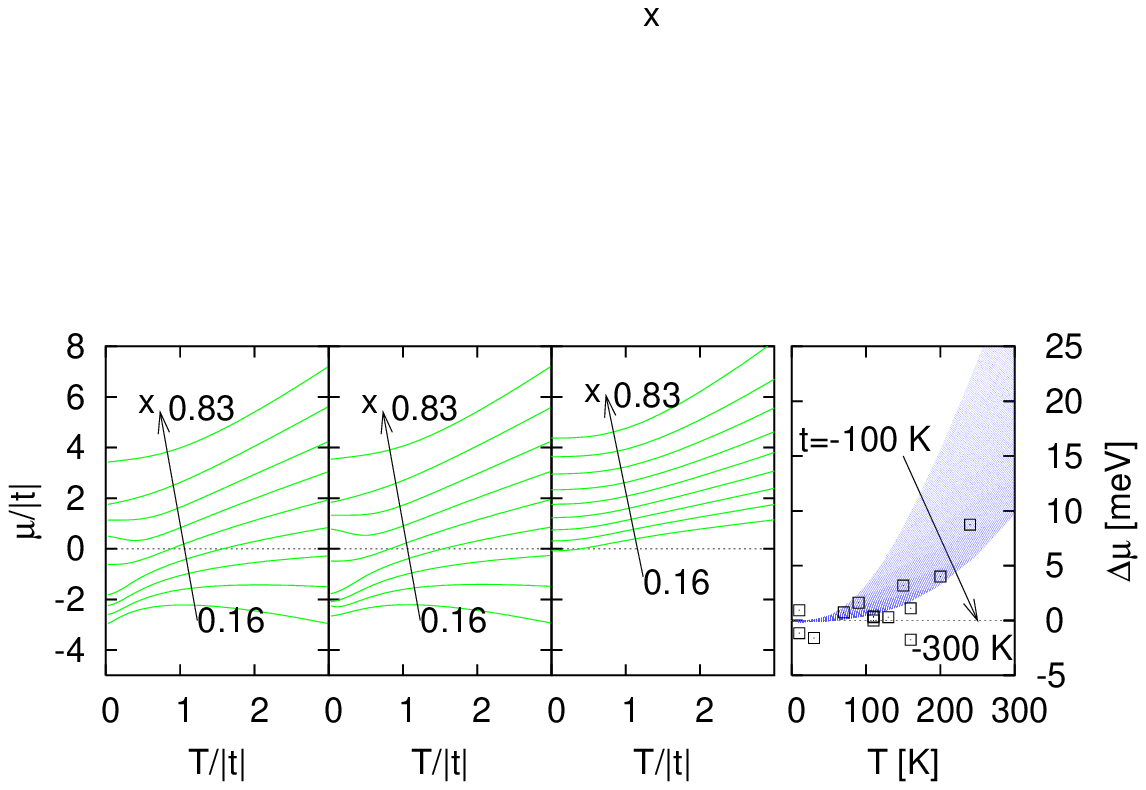}
\caption{(color on-line) Chemical potential $\mu (T)$ for all
accessible dopings on Torus b: $x=2/12=0.16$ to $10/12=0.83$ for $J=0$
(a), $J=0.4|t|$ (b) and $U=0$ (c). Data for $U=0$ computed for
infinite non-interacting system, comparison with (a)-(c) shows
relatively weak effect of interactions in the doping range $x>0.8$.(d)
Squares are experimental data for $\Delta\mu(T)=\mu(T)-\mu(0)$ from
ref. \onlinecite{fujimori}. Shaded area is a comparison with our data
for $U=0$ for $t=-100$ K to $t=-300$ K. }
\label{mu_J5}
\end{figure*}
\end{center}

At larger $x>0.7$ the effect of strong interactions is gradually lost
and the temperature-dependence resembles that of the non-interacting
Fermi-system (Fig. \ref{mu_J5}(c)). For low temperatures well below
the linear range of $\mu$, finite $J>0$ generally increases the value
of the chemical potential with a stronger increase for $x$ near the
Mott-insulator. The high-temperature behavior is independent of $J$.

In Fig. \ref{mu_J5}(d) we compare the results with direct experimental
measurements of the $T$-dependent chemical potential\cite{fujimori} at
$x\approx 0.83$. Since our results in this doping range are rather
insensitive to the value of $U$, we compare with our $U=0$
results. Here, $\mu(T)$ was obtained directly from the tight-binding
density of states by the implicit equation $\langle n\rangle=2\times
\int_{-\infty}^{\infty} f(\epsilon,\mu)\rho(\epsilon)d\epsilon$ with
fixed $\langle n\rangle=1+x$, where $f$ denotes the Fermi-Dirac
distribution function and $\rho$ is the non-interacting density of
states. This fit confirms the notion of NCO as a system with an
unusually small Fermi temperature scale. In Fig. \ref{mu_J5}(d) we
show as a shaded region the values $t=-100$ K to $t=-300$ K,
photoemission results\cite{hasan} point to a value of $|t|\approx
10meV\sim -118$ K, hence $|t|$ here is more than an order of magnitude
smaller than in the cuprate-superconductors. In the following, when
making contact with experiments, we take $t=-100$ K for simplicity.

\subsection{Specific Heat, Entropy}

\begin{center}
\begin{figure*}
\includegraphics[width=6.5cm]{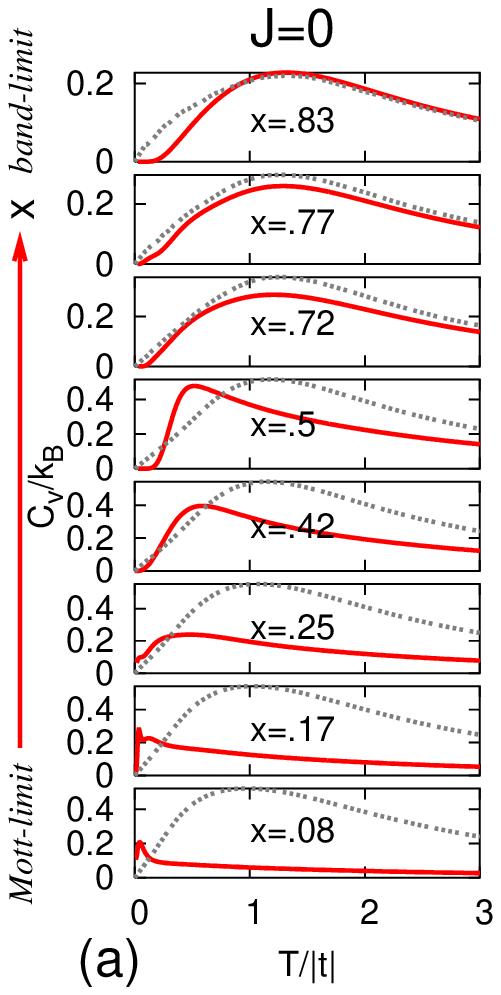}
\includegraphics[width=6.5cm]{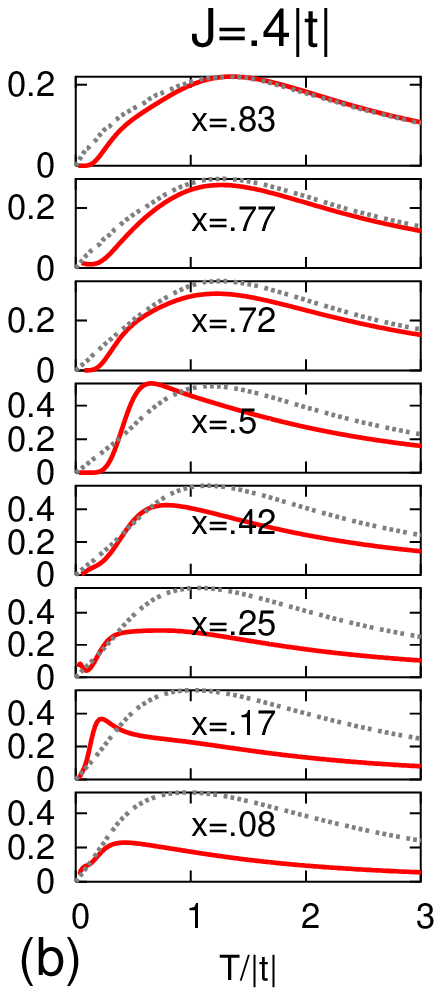}
\caption{(color on-line) Specific heat per site as function of $T$ for
varying $x$ (computed on 18 site ($x\geq 0.72$) and 12 site cluster
($x<0.72$), (a) $J=0$, (b) $J=0.4|t|$. Dotted lines show $C_v(T)$ for
non-interacting fermions and solid line for exact results. Note: scale
changes in plots of different dopings, long temperature exponential
behavior at large $x$ due to finite-system induced gap in energy
spectrum. Results for $x=0.72$ appeared in
ref. \onlinecite{curie_weiss} }
\label{cv_x}
\end{figure*}
\end{center}

\begin{center}
\begin{figure}
\includegraphics[width=8.5cm]{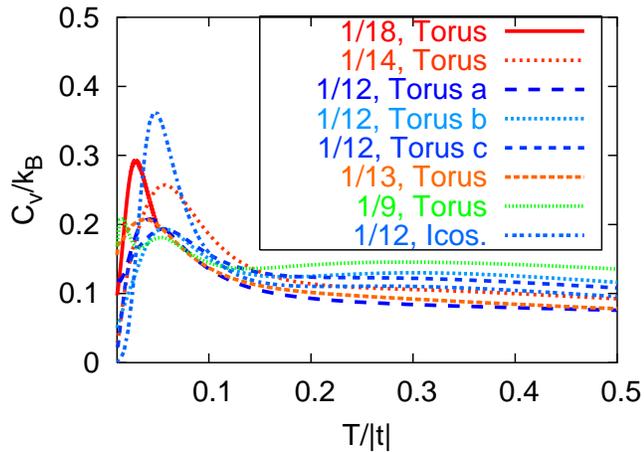}
\caption{(color on-line) $C_v(T)$ at $x=1/L$ for several sets of
boundary conditions, peak in specific heat is system-independent
property. }
\label{cv_antiNag}
\end{figure}
\end{center}

In Fig. \ref{cv_x}(a)-(b) we present the specific heat per site
\[C_v(T)=\frac{1}{L}\frac{d\langle E\rangle }{dT}\]
 with $\langle E\rangle$ the thermodynamic expectation value of the
 internal energy. We first concentrate on the case of $J=0$,
 corresponding to an infinite on-site interaction. Near the band
 insulator ($x\approx 1$) the temperature-dependence resembles that of
 non-interacting fermions, computed at fixed particle density by first
 determining $\mu(T)$ and then the $T$-derivative of the internal
 energy. When the doping is reduced, the main energy scale - manifest
 in a peak of $C_v(T)$ at temperature $T_m(x,J)$  - continuously moves
 to smaller temperatures, away from the tight-binding peak. Even at
 our lowest doping $x=1/L$ the peak-feature persists (called {\it
 kA-peak}). This feature is magnified in Fig. \ref{cv_antiNag} for
 several different systems.

Numerical work on the g.s. of the problem of a single hole in an AFM
spin-background\cite{counter_nagaoka} suggests this system to exhibit
a weak effective AFM interaction $J_{eff}\propto 1/L$ due to the
projected kinetic energy term only. While for one hole this excitation
decays in the thermodynamic limit, $J_{eff}$ nonetheless induces a
N\'eel ordered AFM g.s. and low energy excitations similar to those
observed in the HAF\cite{bernu}.

\begin{center}
\begin{figure}
\includegraphics[width=8.5cm]{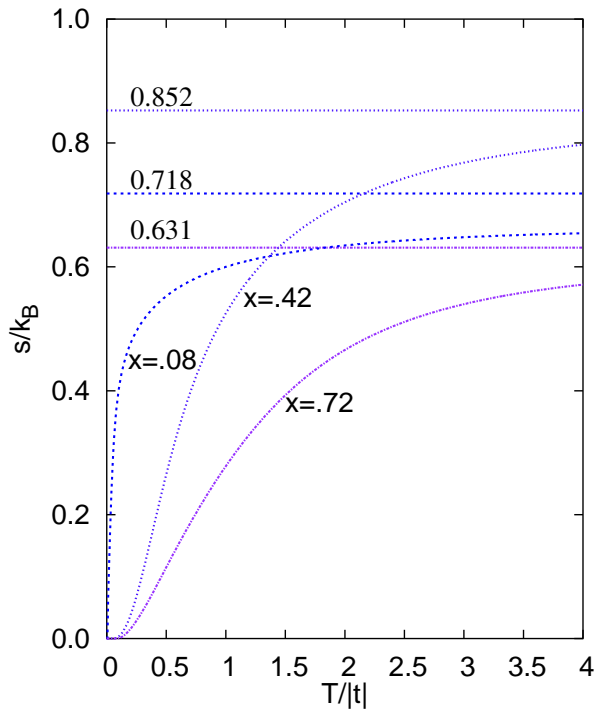}
\caption{(color on-line) Entropy per site for three different dopings
at $J=0$ ($x=1/12=0.08$, $x=5/12=0.42$, $x=13/18=0.72$), horizontal
lines indicate high-temperature limit $s(T\rightarrow\infty)=k_B\log
(g(x))/N$, $x=0.08$ corresponds to Fig. \ref{cv_antiNag}. Exponential
increase at low $T$ for intermediate $x$ due to finite system-size gap
$\Delta$. }
\label{entropy}
\end{figure}
\end{center}

The validity of our computations for the case of a single hole is
underscored by the entropy. In Fig. \ref{entropy}, the temperature
dependence of the entropy shows that the low-energy features observed
in the specific heat are necessary for the entropy to obtain its
(known) high-temperature limit. For $T/|t|\approx 0.075$ half the
entropy is recovered ($s(T)/s(\infty)=0.5$), indicative of
entropy-enhancement due to electronic
frustration\cite{ogata_highT_exp}. For intermediate dopings, the
entropy reaches half its maximum at $T/|t|\approx 0.75$ and for
dopings $x\approx 0.7$ - relevant to the Curie-Weiss phase of NCO - we
find $T/|t|\approx 0.78$.

Within a high-temperature expansion\cite{ogata_highT_exp} a numerical
argument was given for an S-shaped low temperature behavior of the
entropy for the case of $x=0.2$ and $x=0.4$. Our computations at small
dopings rather suggest $s\sim T$ for low $T$ as applicable to a metal
from Fermi-liquid arguments. It is remarkable here that $C_v(T)$
reflects these low energy excitations, notably even for larger dopings
than the one applicable to the Nagaoka-situation.

We now distinguish the effect of $J>0$ (Fig. \ref{cv_x}(b)). This
scenario has been investigated by several authors for the case of the
hole doped square lattice and the
HAF\cite{prelovsek_thermodyn,jaklic_adv_phys}. For the situation of a
single hole $J>0$ shifts the {\it kA}-peak to larger temperatures
$T_m=J/t$. In Fig. \ref{T_star}(b) we present the peak positions as
function of doping and $J$. We find $T_m(x,J)$ to {\it increase}
monotonically both with $x$ and $J$ and the slope as function of $J$
for fixed $x$ to decrease for larger values of $x$. Extrapolating
these results at a given value of $J$ to $x=0$ we obtain roughly
$T_m(x=0)\approx J$. Our result should be distinguished from
investigations on the $t>0$ square lattice\cite{prelovsek_thermodyn},
where $T_m$ was found to follow the opposite trend {\it decreasing}
slightly with $x$ at small dopings, along with a rapid suppression of
the peak. This supports the notion of electronic frustration
contributing to AFM correlations in the present study in opposition to
competing AFM exchange coupling and the FM Nagaoka-mechanism in the
case of $t>0$, capable of quickly eliminating the low-energy spin
excitations.

An experimentally interesting quantity related to the specific heat is
$\gamma$ defined by $C_v(T)\sim \gamma T$ or $\gamma=ds/dT$
corresponding to the low energy electronic part of the specific
heat. When extracting this quantity from our numerical results, we
neglect the contributions at very low $T$ arising purely from the
finite system induced gap in the spectrum (scaling as
$\exp(-\Delta/T)$ where $\Delta$ is the gap between the g.s. and the
first excited state). For the large doping case $x=0.7$ we
found\cite{curie_weiss} $\gamma\approx 45\;(\pm 5)\;mJ/molK^2$
($\gamma_0\approx 33\;mJ/molK^2$ for non-interacting tight-binding
electrons), indicating a significant many-body
enhancement. Experimental results\cite{motohashi} indicate $\gamma
\approx 26\;mJ/molK^2$. On the other hand, in the low-doping regime
the linear increase with $T$ is only observed in a very small
temperature range and this range is possibly better classified as a
non-Fermi-liquid regime. Roughly, $\gamma(x)/\gamma_0(x)\gg 1$ for
$x\gtrsim 0$. The increase in $\gamma$ with decreased doping is also
demonstrated clearly by the low $T$ slope of $s(T)$ in
Fig. \ref{entropy}.

\subsection{Spin Susceptibility}
To gain further insight into the role of spin-fluctuations and to
address the existence of the experimentally observed Curie-Weiss phase
of NCO\cite{suppl}, we compute the spin susceptibility $\chi(T)$ as
function of temperature from the Gibbs free energy $G(T,M)$, here
$M=\sum_i S_i^z$ is the total magnetization. This quantity follows
from the Helmholtz free energy $F(T,H)$ through a
Legendre-transformation $G(T,M)=F(T,M)-H\frac{\partial F}{\partial
H}=F(T,H)+MH$. $H$ is the externally applied magnetic field. \[\chi
^{-1}=\frac{d^2G}{dM^2}\] can then be evaluated numerically, by constraining
the canonical ensemble to the sector of minimal $|S_z^{tot}|$, and
those sectors where $S_z^{tot}$ is raised (lowered) by $\hbar$.

\begin{center}
\begin{figure*}
\includegraphics[height=12cm]{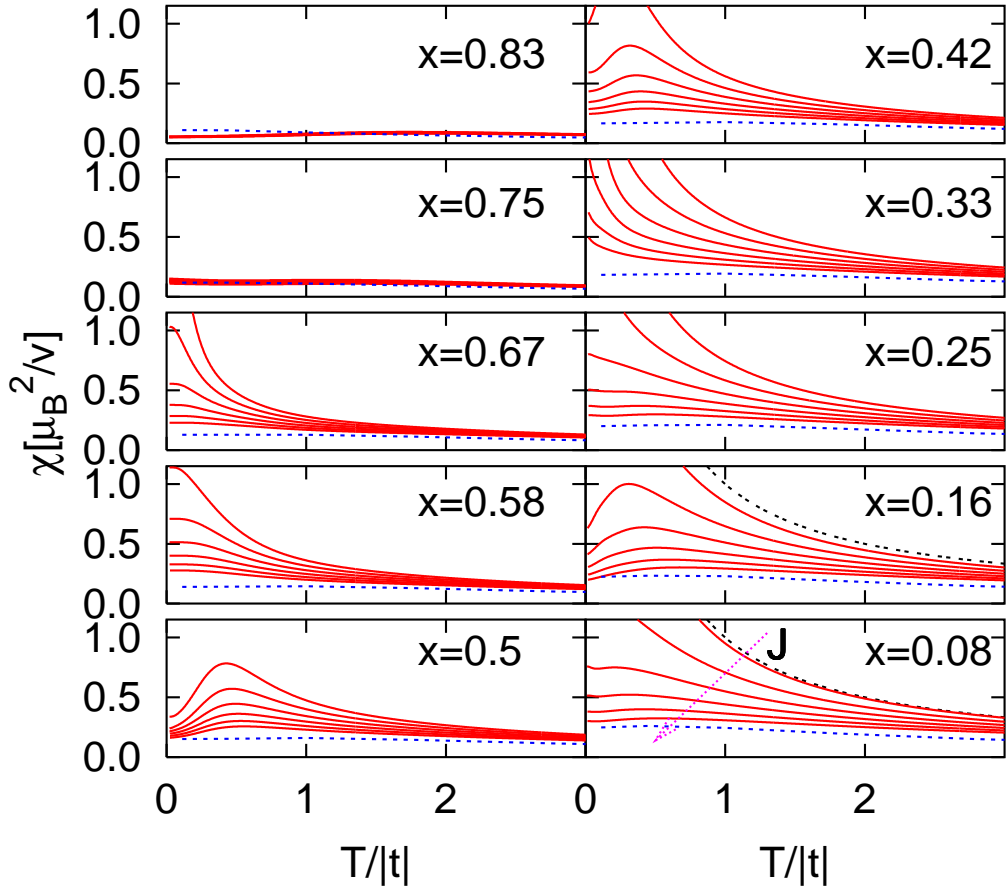}
\caption{(color on-line) Spin susceptibility per site $\chi(T)$ for
all accessible dopings $x$ on 12-site torus. Dependence on
$J/|t|=$($0$, $0.1$, $0.2$, $0.3$, $0.4$, $0.5$), indicated by arrow,
blue dotted line shows bare-susceptibility for non-interacting
fermions of density $n=1+x$ and black dotted line indicates
susceptibility of isolated spin-$\frac{1}{2}$ fermions of density
$n=1$. Plots for $x\geq 0.58$ appeared in
ref. \onlinecite{curie_weiss}.}
\label{CHI}
\end{figure*}
\end{center}
\begin{center}
\begin{figure*}
\begin{tabular}[b]{ccc}
\includegraphics[height=11.5cm]{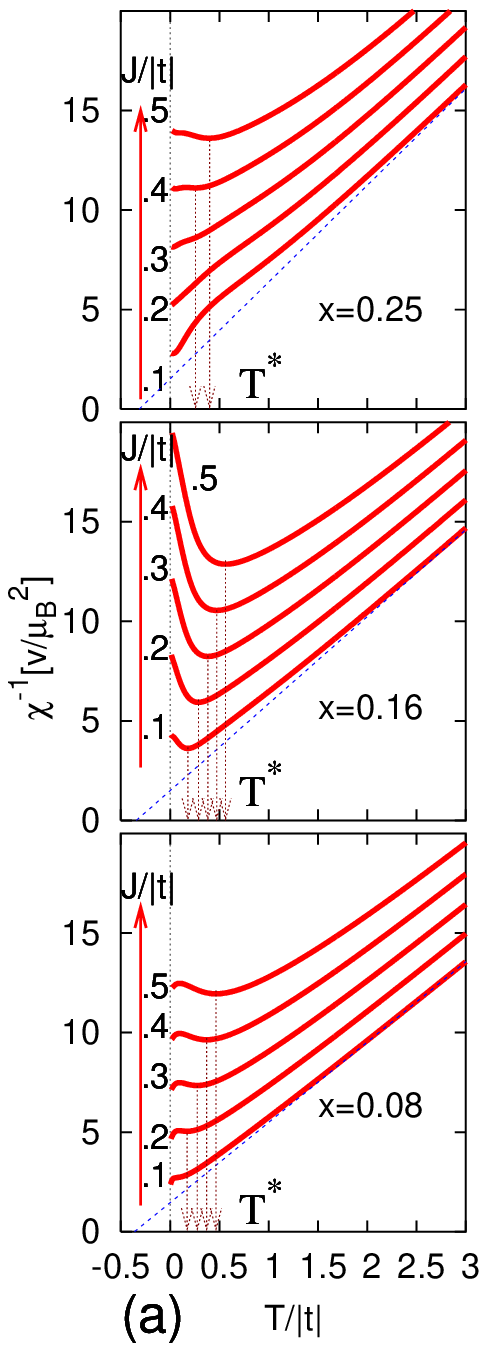}
\includegraphics[height=11.5cm]{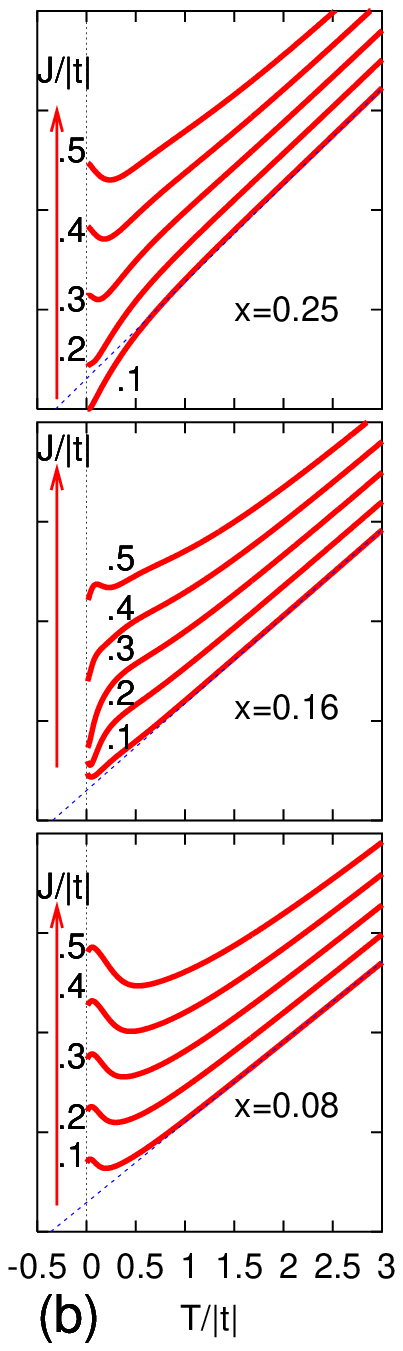}
\includegraphics[height=11.5cm]{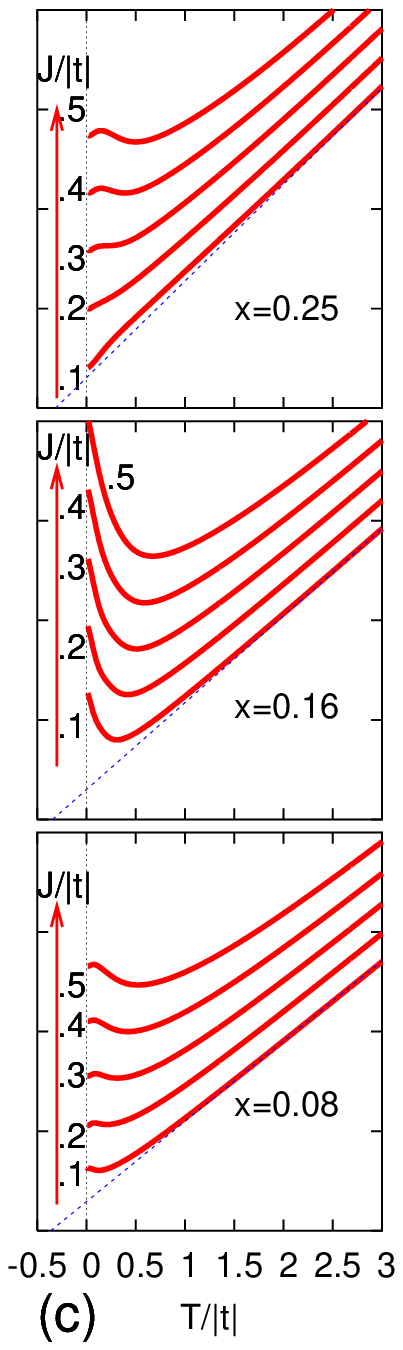}
\begin{tabular}[b]{c}
\begin{picture}(0,0)(0,0)
\put(5,1.5){{\large (d)}} \put(5,-2.0){{\large (e)}}
\put(5,-5.5){{\large (f)}}
\end{picture} 
\includegraphics[width=5cm]{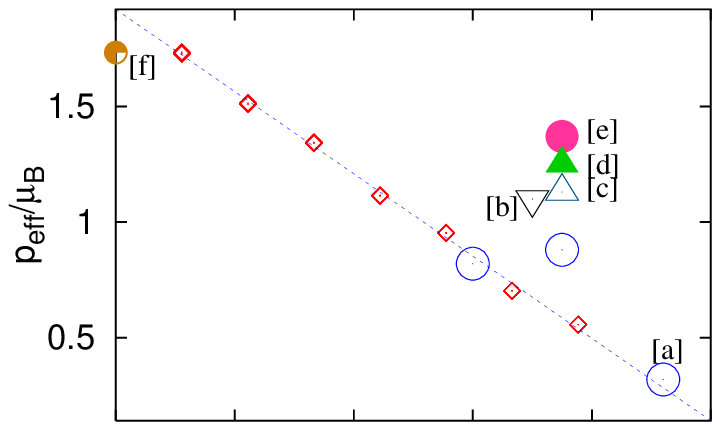}\\
\includegraphics[width=5cm]{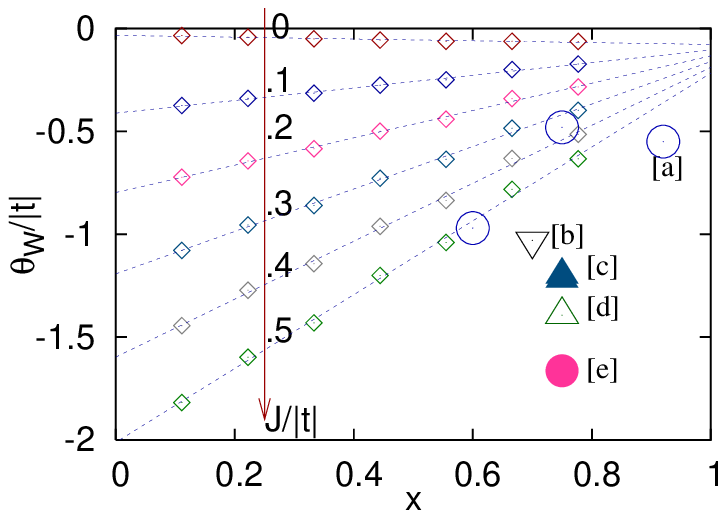}\\
\includegraphics[width=5cm]{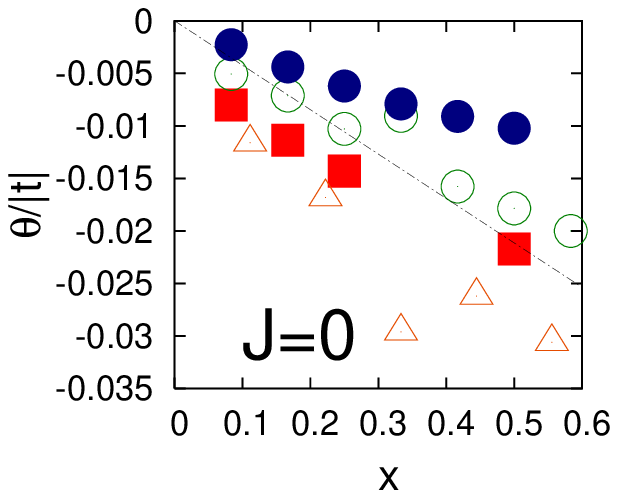}\\
\end{tabular}
&
\end{tabular}
\caption{(color on-line) (a)-(c) Inverse susceptibility
vs. temperature for dopings $x=1/12=0.08$ (bottom), $x=2/12=0.16$
(center), and $x=3/12=0.25$ (top) for Torus a, b and c (l to r). Solid
vertical arrows show increase in $J$, dotted (blue) line shows example
for fit to extract Weiss-temperature $\theta(x,J)$ and effective
moment $p_{eff}$. Dotted vertical arrows indicate temperature $T^*$;
Note: comparison of the three systems allows isolation of finite size
effects, most visible for $x=0.16$. (d) Effective magnetic moment
$p_{eff}$. (e) Weiss-temperature $\theta$ vs doping $x$ for $J=0$
$\ldots$ $0.5$. Letters in angled brackets correspond to experimental
data from [a] Viciu\cite{viciu} et al., [b] Gavilano\cite{gavilano} et
al., [c] Sales\cite{sales} et al., [d] Takeuchi\cite{takeuchi} et al.,
[e] Motohashi\cite{motohashi} et al., and [f] Effective moment of
isolated spin-$\frac{1}{2}$-particles on half-filled lattice. (f)
Magnification of $\theta$ vs. $x$ for $J=0$ for several clusters. }
\label{chi_th_peff}
\end{figure*}
\end{center}

For low temperatures, near the band-limit ($x=1$) we identify weakly
temperature-dependent, Pauli-like behavior of $\chi$, reminiscent of
the experimentally observed spin density wave (SDW)-phase (Fig. \ref{CHI}). When $x$ is
lowered past $x=0.75$ we find\cite{curie_weiss} a crossover to a Curie-Weiss like
behavior with appreciable renormalizations of the g.s. susceptibility
compared to its bare value, mirroring the experimental
findings. This behavior can roughly be understood by
picturing electrons distributed on an empty lattice: Initially, the
many-body eigenstate enables the electrons to move around without
interacting directly with their neighbors, yielding a susceptibility
that is $J$-dependent only to second-order. When the density is
increased ($n>0.25$ or equivalently $n<1.75$), electrons are forced to
interact through the exchange-coupling directly and hence yield a
strongly $J$-dependent magnetic susceptibility. The emergence of the
Curie-Weiss phase is further supported by the strong magnetic field
suppression of the Heikes-Mott term in the thermopower within the
$t$-$J$ model, an effect due to spin-entropy\cite{ong_nature} as was
demonstrated in ref. \onlinecite{curie_weiss}.

For temperatures larger than a certain crossover temperature (which we
call $T^*$), for all dopings, $\chi(T)$ can be described through the
Curie-Weiss form
\[\chi(T)=\frac{1}{3}\frac{L}{V}\frac{\mu_B^2p_{eff}^2}{k_B(T-\theta)}\]
with the Weiss-temperature $\theta$ and the effective magnetic moment
$p_{eff}$, $V$ is the volume of the crystal and $L$ the number of
sites.  For all accessible system sizes - in particular Torus a,b,c
(Fig. \ref{cluster_example}) - we compute $\chi(T)$ to extract
$\theta$, $p_{eff}$ and $T^*$, the parameters characterizing the
magnetic properties at a given doping and interaction strength. To
determine $\theta$ and $p_{eff}$ we fit the high-temperature part of
our computations for $\chi(T)$ to the function $f(T)=a/(T-b)$ (compare
Fig. \ref{chi_th_peff}(a)). In Fig. \ref{chi_th_peff}(a)-(c) we
present the inverse susceptibility at several different dopings near
the Mott-insulating phase. The results for $\theta$ and $p_{eff}$ are
shown in Fig. \ref{chi_th_peff}(d)-(f). For all $x$ and $J\geq 0$,
$\theta$ remains negative indicating AFM correlations. $\theta$ shows
a linear increase when the doping is increased or the interaction is
decreased. In short, as mentioned in ref. \onlinecite{curie_weiss},
$\theta$ can be collapsed into the form $\theta(x,J)=-c\;J_{eff}(x)$
where
\[J_{eff}(x)=J(1+c''x|t|)+c'x|t|\] 
with $c=4.0$, $c'=0.01425$ and $c''=-0.9175$. In contrast to the
relation given by Anderson et al.\cite{NT_weakening} in the case of
the $t>0$ square lattice, in our situation of $t<0$ even at $J=0$
there is a slightly {\it negative} intercept suggesting AFM
spin-interactions in the low temperature regime of the electronically
frustrated $U=\infty$ Hubbard model, indicative of {\it kinetic
Antiferromagnetism (kA)}.

On the other hand, once $J>0$ is allowed, this intercept grows
strongly and expresses dominant AFM correlations
(Fig. \ref{chi_th_peff}(e)). We compare these results with experiments
for various dopings ($x>1/2$) of NCO. Among others, the cases of
$x=0.75$ \cite{motohashi,sales,takeuchi}, $x=0.7$ \cite{gavilano},
$x=0.6$ and $x=0.9$ \cite{sugiyama}, various $x$ in the three-layer
system \cite{viciu} and $x=0.55$ \cite{ando} were experimentally
investigated. The comparison with experiment proves difficult due to
the large number of structural phases \cite{viciu}, the role played by
the Na-ions in the crystal and the large quantitative and qualitative
difference between powder and single crystal samples
\cite{sales}. However, the Weiss-temperature appears to lie between
$100$ K and $200$ K near $x=0.7$ (varying considerably with $H||{\bf
c}$ and $H||{\bf a}$\cite{sales}) and was measured to be smaller at
larger dopings\cite{viciu}. Hence, at least for $x>0.5$, the $t$-$J$
model description yields reasonable experimental agreement.

The effective magnetic moment $p_{eff}$ in units of $\mu_B$ has a
value close to that obtained for localized spin-$\frac{1}{2}$
particles at half-filling $\mu_{1/2}=g\times \sqrt{s(s+1)}=\sqrt{3}$
and decays linearly for $x\rightarrow 1$.

The inverse susceptibility shows a minimum at a temperature $T^*$
which becomes more pronounced with increasing $J$. At low temperatures
the system prefers antiferromagnetic order which is {\it strengthened}
by $J>0$. Thermal spin fluctuations break this order and when $T$ is
increased to values of order $J$ a small perturbative $B$-field is
sufficient to align spins antiparallel to the field. Hence, the
susceptibility is largest at this temperature. When $T$ is further
increased, the ability of the field to flip spins is weakened by
thermal fluctuations, consequently, $\chi(T)$ decreases. The behavior
of this peak as function of hole doping away from the Mott-insulator
has already been investigated within a high-temperature
expansion\cite{ogata_highT_exp}.

\begin{center}

\begin{figure}[h]
\begin{picture}(0,0)(0,0)
\put(-.8, 3.5){{\large (a)}} \put(-.8, -2.5){{\large (b)}}
\end{picture} 
\includegraphics[width=6.5cm]{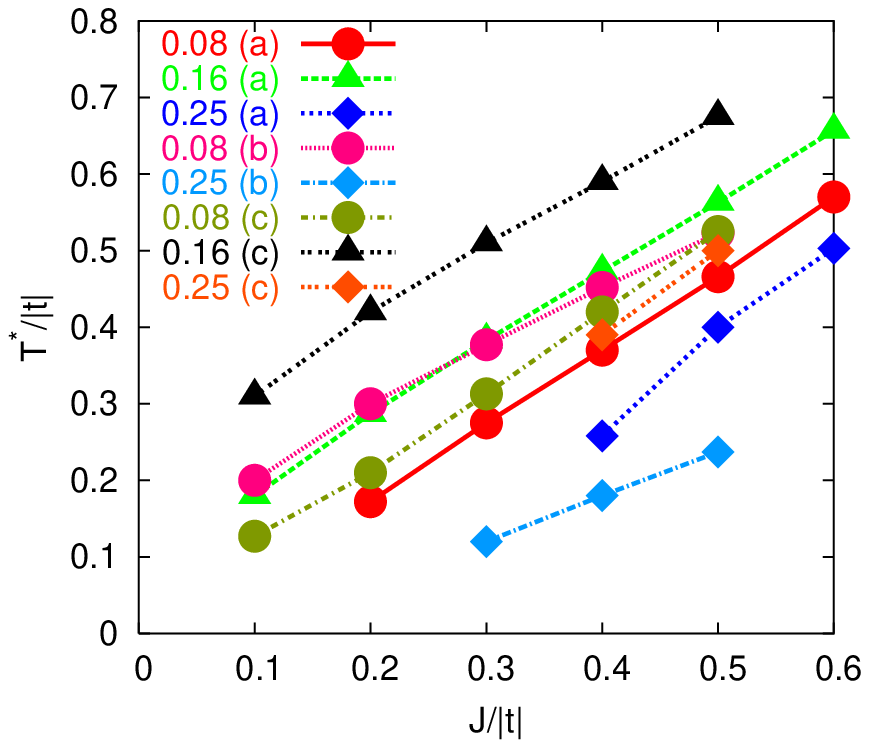}\\
\includegraphics[width=6.5cm]{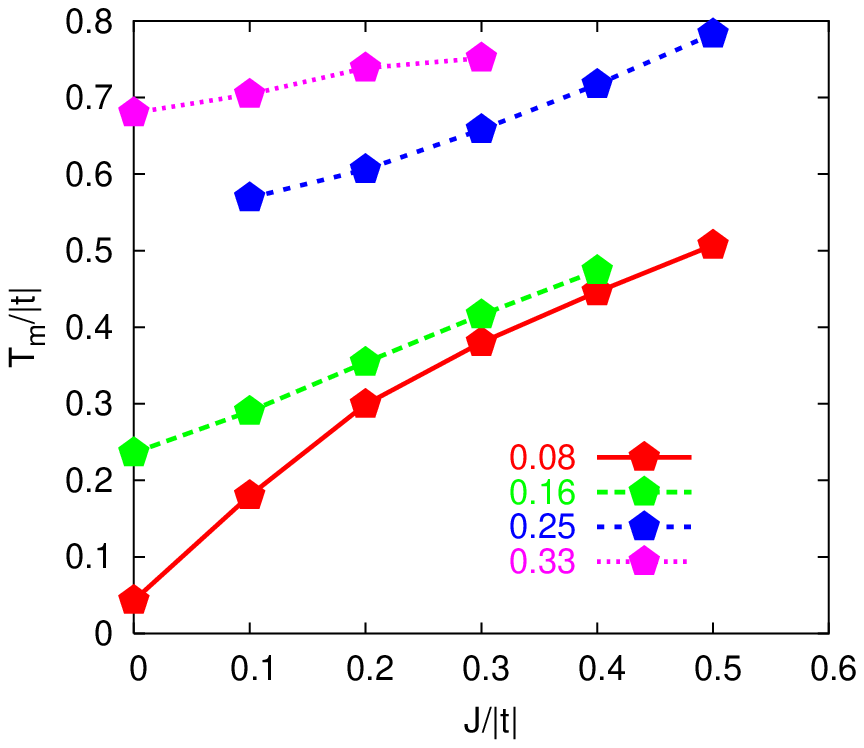}
\caption{(color on-line) (a) AFM-paramagnetic crossover-temperature as
function of $J$ for several dopings on 12-site torus clusters a, b,
c. (b) Evolution of $T_m$ as function of $J$, values averaged over
three 12-site clusters a, b, c. }
\label{T_star}
\end{figure}
\end{center}

Due to the small size of the systems available to our study, it is
difficult to confirm the results for the doping
dependence\cite{ogata_highT_exp}. Here, we study the behavior as
function of $J$, considering several dopings near half-filling and
several system geometries (Fig. \ref{T_star}(a)). The figure indicates
that $T^*/|t|$ roughly scales as $J/|t|$, hence, the scale at which
the system undergoes a transition from an AFM-ordered to a
paramagnetic state is set by the interaction strength. As was shown
for the case of $J=0$ and $t>0$ in the hole doped
system\cite{counter_nagaoka}, a single hole favors AFM N\'eel order
and numerical results\cite{bernu} indicate that the triangular lattice
Heisenberg model also favors N\'eel order - at least in the g.s. and
the elementary excitations - hence it is not surprising that a finite
$J>0$ emphasizes this tendency. This is in opposition to the case of
$t<0$ where the kinetic energy term favors ferromagnetism and competes
with the Heisenberg-interaction preferring AFM.

\subsection{Hall Coefficient}
Experimentally, NCO has been shown to exhibit unusual thermoelectric
properties, resulting from strong electron correlations. Wang et
al.\cite{wang_hall} measured the dc-Hall coefficient as function of
temperature at several dopings showing unexpected temperature
dependence with an oscillatory behavior below $200$ K and a rapid
linear increase for temperatures above $200$ K.

Much theoretical work has been invested into the study of $R_H$ on
square
lattices\cite{sss,kumar,castillo,prelovsek,prelovsek_D,prelovsek_T,prelovsek_reactive,lange,rojo,assaad,mahesh,ong,motrunich,anderson_hall_angle,phillips}
both for g.s. properties and for finite temperatures. For the
triangular lattice, it was possible to predict the high-temperature
behavior within a high-temperature expansion of the
$t$-$J$-model\cite{sss,kumar} in its high-frequency limit. Several
authors have also attempted to describe this behavior through
Boltzmann transport theory\cite{motrunich,ong} within the relaxation
time approximation, where $R_H$ is expressed as an integral over the
Fermi-surface curvature and the relaxation time $\tau_{\bf {\it k}}$
is taken to be isotropic. For large dopings - $x\geq 0.7$ - where the
hole-like Fermi-surface of NCO approximates a circle, this leads to
the same high-temperature linear coefficient as the
$t$-$J$-high-temperature expansion, however, the behavior completely
diverges in the dense regime ($x<0.7$), possibly due to the misleading
replacement of the interacting Fermi-surface by its non-interacting
counterpart. Also, at low temperatures - where the relaxation time is
anisotropic - the oscillatory behavior seen in experiments is not
reproduced.

To shed some light on the low-temperature and strongly-correlated
behavior, we start from the high-frequency limit\cite{sss} for
$R_H(T,\omega)$:

\begin{eqnarray}
R_H^*(T) & \equiv & \lim_{\omega\rightarrow\infty}R_H(T,\omega)
\notag\\\label{RH_eq} & = & \lim_{B\rightarrow 0}\left(-\frac{i v
L}{Be^2}\frac{<[J_x,J_y]>}{<\tau_{x,x}>^2}\right)
\end{eqnarray}

Here, $e$ is the electron charge, $J_x$ ($J_y$) are the currents in
$x$ ($y$)-directions, $v$ is the unit cell volume, and $\tau_{x,x}$ is the diagonal part of the
stress tensor. To ensure the effect of strong correlations within this
limit, it is necessary to force the hierarchy $t<\omega \ll U$. We
enable this by projecting out double occupancy using the Gutzwiller
projector, hence the $t$-$J$ model is the appropriate description.

Our plan is as follows: we show the connection of this high-frequency
limit - not measurable in experiments - to the dc or low-frequency
behavior. Then we evaluate the dependence on the parameters $T$, $J$
and $x$ for several systems. Hence, for the first part we compare
$R_H^*$ with the expression derived from the Kubo formulae for the
electrical conductivities
\begin{widetext}
\begin{equation}\sigma_{\alpha,\beta}=\frac{ie^2}{\omega v}\left(<\tau_{\alpha,\beta}>-\frac{1}{Z}\sum_{\mu,\nu}\frac{\exp -(\beta\epsilon_{\nu})-\exp -(\beta\epsilon_{\mu})}{\epsilon_{\mu}-\epsilon_{\nu}-\omega-i\eta}<\nu|J_{\alpha}|\mu><\mu|J_{\beta}|\nu>\right)
\;,\end{equation}
\end{widetext}
with $\epsilon_{\nu}$ the $\nu$'th energy eigenvalue of $\hat{H}$ and
$Z$ the canonical partition function. In terms of this expression the
Hall coefficient is given by \beq
R_H(\omega)=v\frac{\partial}{\partial
B}\left(\frac{\sigma_{x,y}(\omega)}{\sigma_{x,x}(\omega)\sigma_{y,y}(\omega)}\right)_{B=0}.\eeq

As a transport object $R_H$ is a demanding quantity to study on finite
clusters. The effect of boundary conditions and their finite size
implications are difficult to eliminate. We introduce a small magnetic
field through the Peierls substitution (Eq.~\ref{peierls}) which
impacts only on the orbital motion of the electrons and does not
couple to the spin. Studying $R_H$ on toroidal and spherical
geometries reduces the finite size effects on the transport, however,
these periodic systems constrain the smallest accessible flux $\phi_0$
to $\pi/L$. For $L=12$ this is already a considerable flux. An
alternative are ladder systems such as in
Fig. \ref{cluster_example}(b) where the limit $\phi_0\rightarrow 0$ is
possible. However, ladders have additional finite size effects
impacting on the transport quantities.

\begin{center}
\begin{figure}[h]
\includegraphics[width=8.5cm]{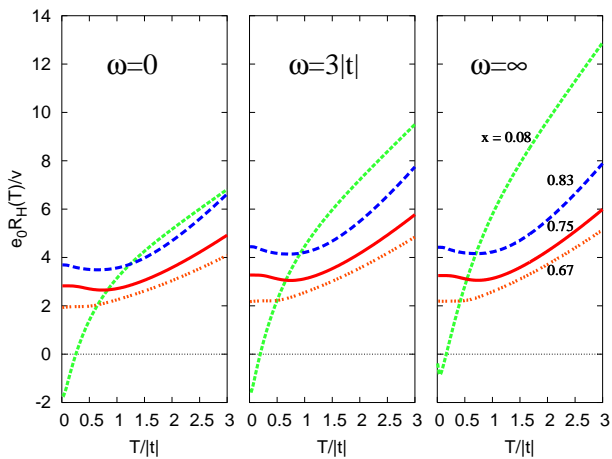}
  \caption{(color on-line) $R_H(\omega,T)$ on 12-site torus at several
  frequencies for several dopings and $J=0$. }
\label{RH_T_w}
\end{figure}
\end{center}

For several dopings ($x=0.08$, $0.67$, $0.75$, $0.83$) we compute
$R_H(\omega)$ as a function of temperature both on the torus and the
icosahedron. To take into account the finite size of the system, a
broadening $\eta$ of the energy equal to the mean level spacing of the
current matrix elements was employed, this requires $\eta\approx
3|t|$. In Fig. \ref{RH_T_w} we show $R_H$ as function of temperature
at three different frequencies. Comparing $\omega=0$ to
$\omega=\infty$ we find that the dc-limit is qualitatively the same as
the high-frequency limit, however, the overall temperature-dependence
is scaled by a renormalization factor $s\equiv
R_H(T,\omega=0)/R_H(T,\infty)$ where $s$ is only weakly temperature
dependent. For the torus, $s$ was found to be $0.83$ ($0.78$) at
$x=0.83$ ($x=0.75$). At $x=0.083$ the many-body renormalization is
stronger.

\begin{center}
\begin{figure}[h]
\includegraphics[width=8.5cm]{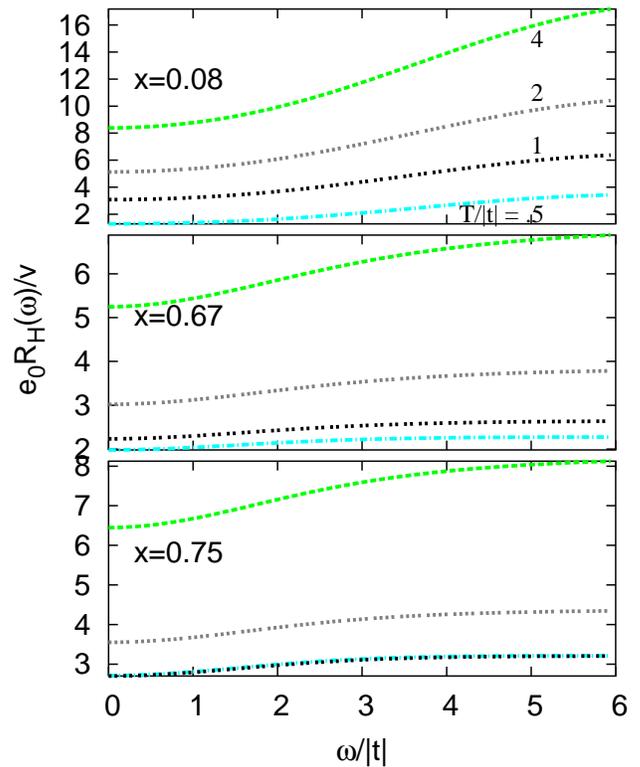}
  \caption{(color on-line) $R_H(\omega)$ at several temperatures and
  dopings as function of frequency on 12-site Torus a for
  $J=0$. Temperatures in all plots $T/|t|=(0.5,1,2,4)$. }
\label{RH_w_x}
\end{figure}
\end{center}

Conversely, in Fig. \ref{RH_w_x} we show the frequency-dependence of
$R_H$ at several temperatures and different dopings. Similar to the
work done by Grayson et al.\cite{drew} on the Hall angle, finite
frequency experiments may be helpful to confirm the weak
frequency-dependence seen in our computations: With $|t|\approx 10meV$
results in the frequency range $\omega\geq 3|t|\approx 30meV$ could be
compared to our theoretical results and hence lend experimental
support for the usefulness of the high-frequency limit\cite{sss} of
$R_H$.

We have computed two additional quantities related to the dc-limit as
checks of the validity of the dc-computations. Zotos et
al.\cite{prelovsek_D} derived an expression for the g.s. expectation
value, $R_H(T=0)=\frac{1}{|e|}\frac{d}{dn}\log D(n)$ with $D$ the
charge stiffness $D=\frac{d^2\langle E\rangle _0}{d\phi^2}$, $\langle
E\rangle _0$ the g.s. energy and $\phi$ a constant flux inducing a
persistent current\cite{twisted_bc}. In this fashion, we evaluated the
doping dependence of the g.s. and found $R_H^0>0$ in the experimental
regime near $x\approx 0.7$ and a divergence $R_H^0\rightarrow -\infty$
as the Mott-limit is approached\cite{curie_weiss}. The second check
refers back to the basic definition of the dc-Hall-coefficient
$R_H=-E_y/(Bj_x)$ and computing this directly on a ladder system. This
computation showed no change of sign for $x\approx 0.7$ as function of
$T$ and a qualitatively similar behavior to the $\omega=\infty$
result.


Encouraged by the notion of weak frequency-dependence, we studied
$R_H^*\equiv R_H(\omega=\infty)$. We have studied several 12 site
tori, several doping values on an 18 site torus, the icosahedron and
several ladders.

The g.s. $R_H^*(x)$ shows a Drude drop-off near the band insulator and
a saddle point around $x=0.5$ as was reported by the authors in
ref. \onlinecite{curie_weiss}. When the strong-correlation regime near
the Mott-insulator is approached ($x<1/2)$, the slope again increases
and for 1 hole ($x=0.083$) $R_H^0$ changes sign.

\begin{center}
\begin{figure}[h]
\includegraphics[width=4cm]{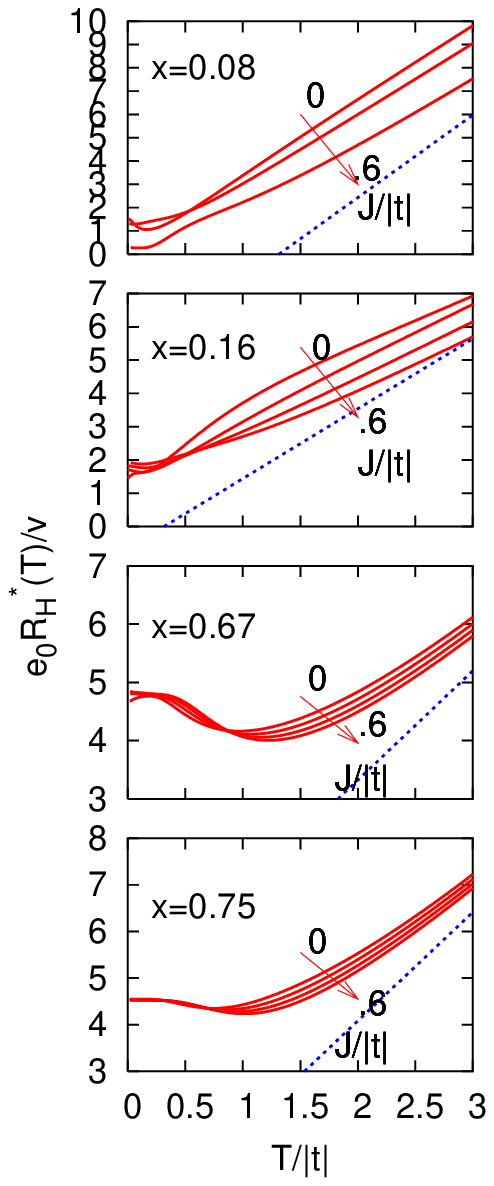}
\includegraphics[width=4cm]{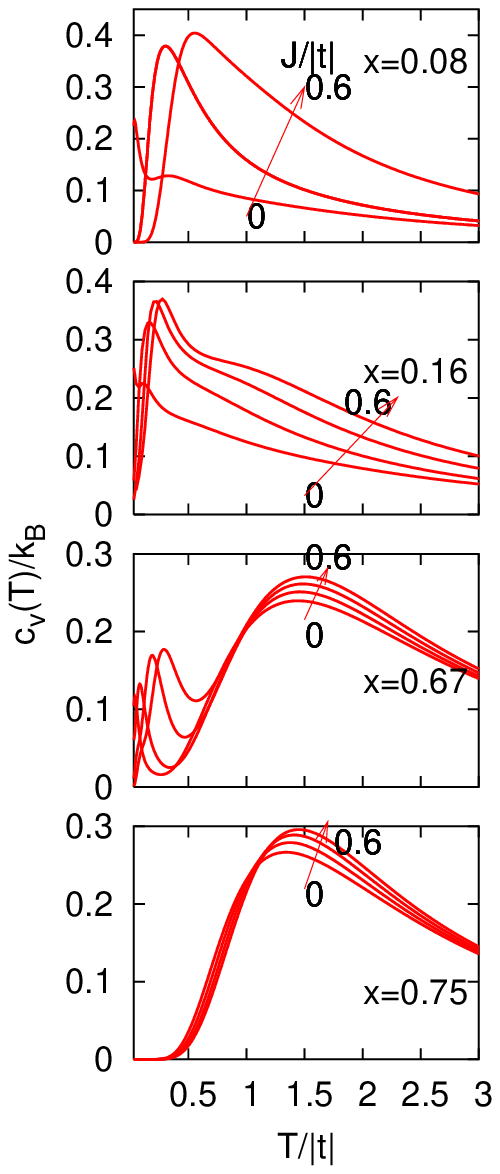}
  \caption{(color on-line) (left) $R_H^*(T)$ for several $x$ and $J$
  for Torus b. (right) $C_v(T)$ for several $x$ and $J$ on same Torus.}
\label{RH_T_J}
\end{figure}
\end{center}

The high temperature behavior confirms the prediction from the
high-temperature series expansion\cite{sss,kumar}. The interesting
temperature range is for $T<2|t|$ where the experimental data shows an
oscillatory behavior and analytical results are not available
(Fig. \ref{RH_T_J}). Our computations show a distinct minimum in
$R_H(T)$ which appears near $T=|t|$ in the Curie-Weiss regime
($x\approx 0.7$) and continuously moves to smaller temperatures as the
doping is decreased. For the smallest dopings accessible ($x=1/L$)
this minimum appears near $T=0.2|t|$, while the behavior becomes
linear at higher $T$ in all cases. In the Curie-Weiss regime some
systems also show a weak maximum close to $T=0.25|t|$. These results
point towards the existence of two distinct energy scales. The
pronounced minimum appears to be correlated with the peak in the
electronic specific heat (Fig. \ref{RH_T_J}), which also occurs around
$T=|t|$ in the band limit, and gradually shifts towards smaller
temperatures with decreased doping.

To strengthen this point it is useful to include the dependence on
$J$. As reported for the specific heat, the effect of $J$ is to move
the peak in specific heat to larger temperatures. This behavior is
mirrored by $R_H$. In Fig. \ref{RH_T_J} we show $R_H(T)$ for several
dopings and $J$. At $x=0.75$ the peak of $C_v(T)$ shifts from
$T=1.3|t|$ to $T=1.5|t|$ while the minimum in $R_H(T)$ moves from
$T=0.75|t|$ to $T=1.1|t|$. At $x=0.67$ the behavior around $T=|t|$ is
similar, however an additional peak in the specific heat emerges at
$T<0.5|t|$. A qualitatively similar feature is seen in other clusters,
however, its magnitude is dependent on the system geometry. Hence, we
do not attempt to extract {\it quantitative} information in the
$T$-range below $T=0.5|t|$. Nonetheless, it is important to note that
with decreasing doping a second energy scale - caused by spin
interactions - becomes dominant. At small doping $x=0.16$ and $x=0.08$
the specific heat becomes strongly $J$-dependent and the main peak in
the specific heat - as discussed in the first section - increases
significantly with $J$. The variations in $R_H(T)$ with $J$ are also
more pronounced.

\begin{center}
\begin{figure}[h]
\includegraphics[width=8.5cm]{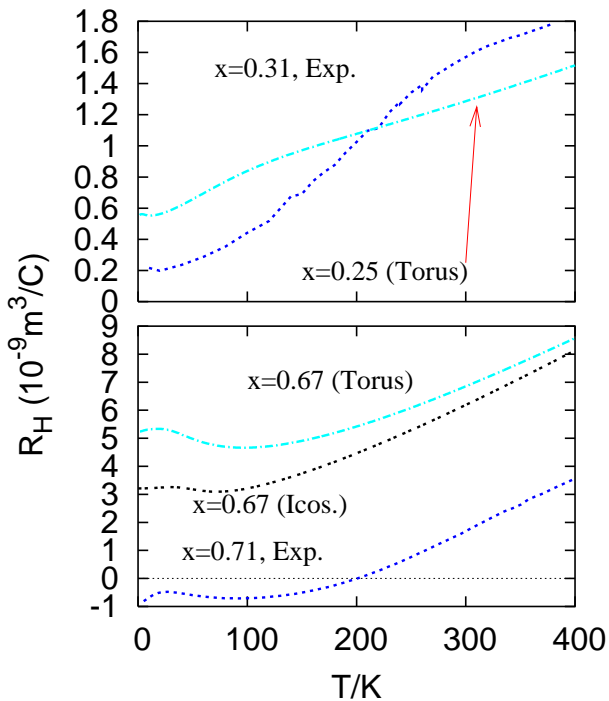}
  \caption{(color on-line) Comparison of 12-site Torus b and
  icosahedron with experiment\cite{ong} at $x=0.31$ and $x=0.71$, we
  take $t=-100$ K, in the lower plot our numerical results are scaled
  to match the slope of the linear part at $T>200$ K in the
  experimental results.}
\label{ong_31}
\end{figure}
\end{center}

Assuming $t\approx -100$ K we compare our results to the
experimentally available data\cite{wang_hall} at $x=0.31$ and
$x=0.71$, this was done in ref. \onlinecite{curie_weiss} for
$x=0.71$. In Fig. \ref{ong_31} the comparison at $x=0.31$ shows good
agreement for the slope at high $T$ and reproduces the temperature
scale for the minimum near $25$ K. For $x=0.71$ the temperature scale
for the minimum near $100$ K is seen both in experiment and
theory. The slope of the high temperature limit differs by a factor of
4 and was adjusted in the plot for better comparison. Also, the theory
does not show the change of sign as function of temperature. However,
the oscillatory behavior below $T=100$ K and the onset of the linear
increase for $T>200$ K agrees well with experiment. Hence, the exact
temperature dependence of $R_H$ remains a challenging task - both on
the theoretical and experimental side.


\subsection{Diamagnetic Susceptibility}
\begin{center}
\begin{figure}
\begin{picture}(0,0)(0,0)
\put(-0.5, 3.5){{\large (a)}} \put(-0.5, -3.){{\large (b)}}
\end{picture} 
\includegraphics[width=7cm]{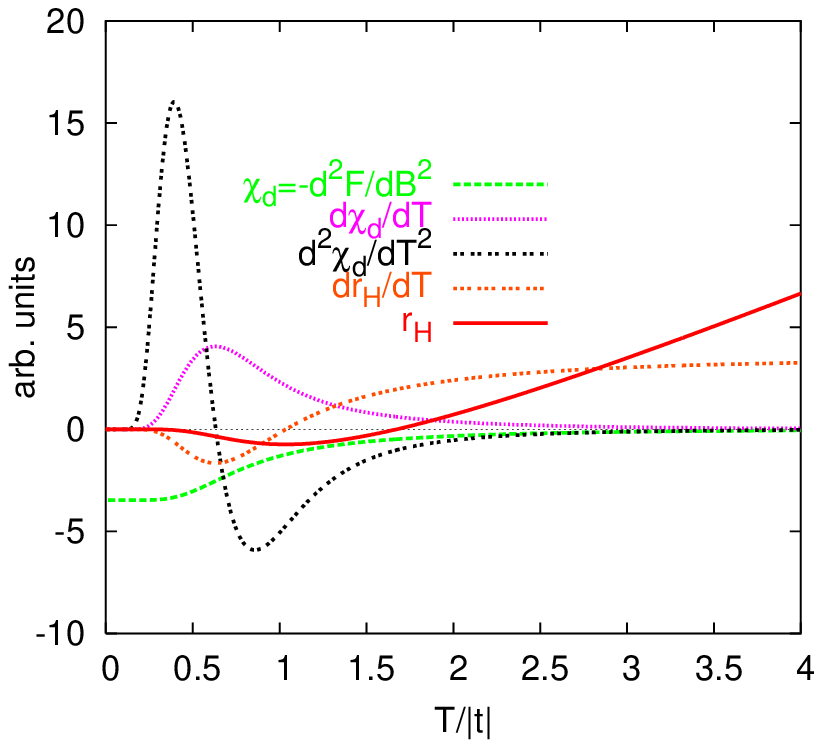}
\includegraphics[width=7cm]{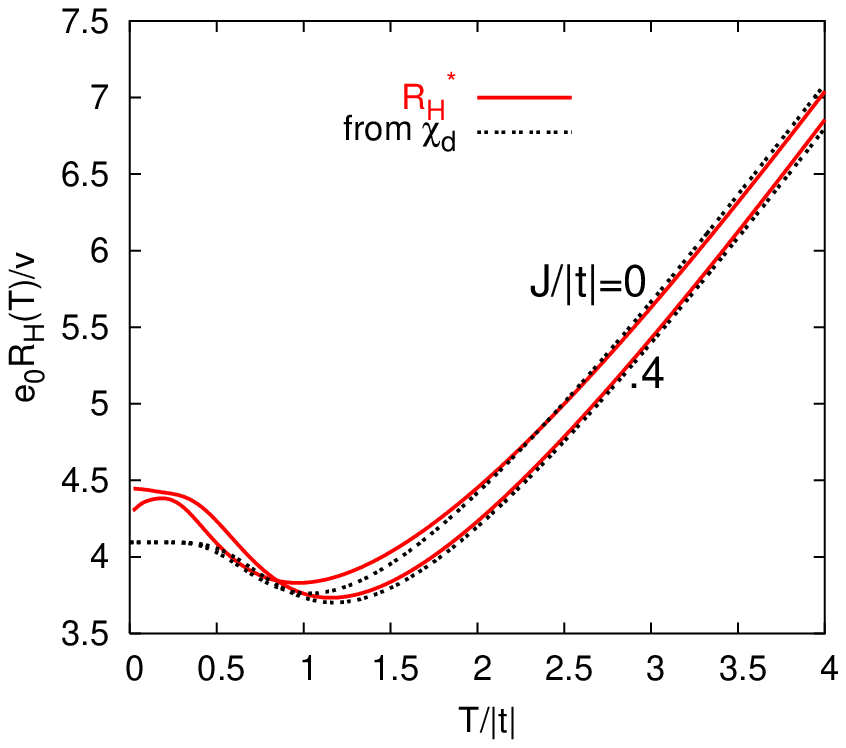}
\caption{(color on-line) (a) $\chi_d(T)$ and unitless Hall-coefficient
$r_H$ and their derivatives on a 12-site ladder at $x=0.83$. (b)
Comparison of numerical results for $x=0.83$ for $R_H^*(T)$ (obtained
on an 18-site torus) with the integral of Eqtn. (\ref{chiL_RH_eq})
($\chi_d(T)$ obtained on a 12-site ladder system) for two different
values of $J$. Appeared partly in ref. \onlinecite{curie_weiss}.}
\label{chiL}
\end{figure}
\end{center}

Lastly, we study the orbital susceptibility per site \[\chi_d(T)\equiv
-\frac{1}{L}\frac{\partial^2F}{\partial B^2}\;,\] where $F$ is the
free energy and $B$ the magnitude of the perpendicular magnetic
field. We define the dimensionless flux $\alpha=BA/\phi_0$ through a
triangular plaquette with $A=\sqrt{3}a^2/4$ the area of the plaquette
and $a$ the lattice spacing. We get
\[\chi_d=-\frac{1}{L}\frac{\partial^2 F}{\partial\alpha^2}\left(\frac{\partial\alpha}{\partial B}\right)^2\;.\] 
It is useful to employ ladder systems as shown in
Fig. \ref{cluster_example}(b). Defining $t=\hbar^2/2ml^2$ we have
\[\frac{1}{L}\chi_d=-\frac{2m}{\hbar^2l^2}A^2\mu_B^2\frac{1}{L}\frac{\partial f}{\partial \alpha}\;\;,\]
where $f$ is the dimensionless $F/t$ and $\mu_B$ is the
Bohr-magneton. In Fig. \ref{chiL}(a) we present the diamagnetic
susceptibility and its first two derivatives with respect to $T$. For
free electrons in an infinite system, $\chi_d$ corresponds to
electrons moving in circular orbits with radii inversely proportional
to the magnetic field. $R_H$ measures the polarization induced in a
system with a stationary current and perpendicular magnetic field. It
is hence reasonable to ask for a connection between these two
quantities. Our numerics suggest the relation
\begin{equation}\label{chiL_RH_eq}T\frac{\partial^2\chi_d}{dT^2}=c(x)\;\frac{\partial^2 R_H}{\partial T^2}\end{equation} where $c(x)$ is a function depending on doping $x$. 

In Fig. \ref{chiL}(b) we display our numerical result for $R_H^*(T)$
at a doping relevant to the experimental situation\cite{wang_hall}
compared to the result obtained from integrating
Eqtn. (\ref{chiL_RH_eq}) and choosing the two integration constants
appropriately.  For $J=0$ this was done by the authors in ref. \onlinecite{curie_weiss}.  
Ignoring the range for $T<0.5|t|$ which may be strongly
influenced by finite size effects, we find that both curves show a
minimum at $T\approx |t|$ and then increase until $T=2|t|$. The
reaction to finite $J>0$ is also similar for both curves, moving the
minimum from $T=|t|$ to roughly $T=1.25|t|$. We plan to further
investigate this connection within a high-temperature expansion for
$\chi_d(T)$, in particular for the doping regime close to the
Mott-insulator.

\section{Conclusion}
In summary, we have presented a systematic study of the entire doping
regime for the triangular lattice $t$-$J$ model with electronic
frustration ($t<0$ for $n>1$). We have identified low energy
excitations, stemming purely from kinetic processes ($J=0$) of
infinitely correlated spin-$\frac{1}{2}$ fermions. These are manifest
in a peak structure in the specific heat emerging at very low $T$ near
the Mott-insulator, however persisting for intermediate dopings. The
effect of electronic frustration is further supported by
entropy-enhancement in the underdoped regime with a low entropy
recovery temperature.

We have studied both the diamagnetic and spin-susceptibility.  For the
spin susceptibility we extract a relation for effective
spin-interactions $\theta(x,J)$ valid for all dopings. This stands in
contrast to Anderson's formula for FM $t>0$, since in our case AFM is
{\it supported} rather than suppressed by doping. In the
experimentally relevant doping range of $x>0.5$ we offer an
explanation for the crossover between the Curie-Weiss and SDW metallic
phases, as observed in experiments\cite{foo}. For the diamagnetic
susceptibility $\chi_d$ a connection is obtained between $\chi_d$ and
the temperature dependent Hall coefficient.

The dynamic Hall coefficient reveals a weak frequency-dependence and a
qualitatively similar dc and high-frequency behavior, with stronger
renormalizations close to the Mott-limit. This strengthens the
practical relevance of the high-frequency limit\cite{sss} $R_H^*$. The
temperature-dependence confirms the linear slope at high
temperatures\cite{sss} and reproduces the experimentally observed
minimum near $T=100$ K at $x=0.71$.

In conclusion, the $t$-$J$ model is applicable to NCO for $x>0.5$ and
shows that the interface of strong correlations and metallic behavior
indeed describes the experimental situation here. For $x<0.5$ the
frustrated $t$-$J$ model leads to subtle effects of strong
correlations, even in the case of $J=0$. Our results show, that for
the proper description of experimentally observed normal state
properties\cite{foo} there - such as the charge-ordered insulator at
$x=0.5$ or the paramagnetic metal for $x<0.5$ - an extension of the
simple $t$-$J$ model would be necessary.

\begin{acknowledgments}

This work is supported by Grant No. NSF-DMR0408247.

\end{acknowledgments}

\end{document}